\documentstyle [12pt,a4p,epsfig,amsmath,multicol]{article}
\begin{document}
\vspace*{0.6cm}

\begin{center} 
{\normalsize\bf Space-time symmetry is broken}
\end{center}
\vspace*{0.6cm}
\centerline{\footnotesize J.H.Field}
\baselineskip=13pt
\centerline{\footnotesize\it D\'{e}partement de Physique Nucl\'{e}aire et 
 Corpusculaire, Universit\'{e} de Gen\`{e}ve}
\baselineskip=12pt
\centerline{\footnotesize\it 24, quai Ernest-Ansermet CH-1211Gen\`{e}ve 4. }
\centerline{\footnotesize E-mail: john.field@cern.ch}
\baselineskip=13pt
\vspace*{0.9cm}
\abstract{Space-time intervals corresponding to different events on the worldline
 of any ponderable object (for example a clock) are time-like. In consequence, in
 the analysis of any space-time experiment involving clocks only the region for $c\Delta t \ge 0$
 between the line $\Delta x = 0$ and the light cone projection $c\Delta t = \Delta x$ of the $c\Delta t$ 
 versus $\Delta x$ Minkowski plot is physically relevant. This breaks the manifest space-time symmetry
 of the plot. A further consequence is the unphysical nature of the `relativity of simultaneity'
 and `length contraction' effects of conventional special relativity theory. The only modification
 of space-time transformation laws in passing from Galilean to special relativity is then
 the replacement of universal Newtonian time by a universal (position independent) time dilation
 effect for moving clocks.}  
\vspace*{0.9cm}
\normalsize\baselineskip=15pt
\setcounter{footnote}{0}
\renewcommand{\thefootnote}{\alph{footnote}}

 The concept of spontaneously broken symmetry is a ubiquitous one in modern physics.
 Originating in solid-state theory~\cite{Anderson,Nambu}, it is the basis of
 the Higgs mechanism of the standard model of particle physics~\cite{WeinQFT}. 
 As exemplified by the behaviour of a ferromagnet, spontaneous symmetry breaking 
 occurs when the fundamental laws of some physical phenomenon respect a certain
 symmetry (rotational invariance in the case of a ferromagnet) which is broken in
 an actual physical realisation of the phenomenon. The fundamental laws are encapsulated
 in a Hamiltonian in non relativistic quantum mechanics, by a Lagrangian in relativistic
 quantum field theory, and by differential equations, such as Newton's Second Law of mechanics,
 or Maxwell's equations, in classical physics.
\par The fundamental laws of special relativity theory (SRT) are also encapsulated in
   differential equations, the Lorentz transformations (LT) for space and time intervals:
   \begin{eqnarray}
   \Delta x'& = & \gamma(\Delta x -\beta\Delta x^0), \\
     \Delta (x^0)'& = & \gamma(\Delta x^0 -\beta\Delta x)  
  \end{eqnarray}
   where $x^0 \equiv ct$, $(x^0)' \equiv ct'$, $\Delta x \equiv x_1-x_2$ etc, $\beta \equiv v/c$,
   $\gamma \equiv 1/\sqrt{1-\beta^2}$ and $c$ is the speed of light in free space.
    The parallel $x$ and $x'$ coordinate axes are defined
   in the inertial frames S and S' respectively. The frame S' moves with speed $v$ in the
   direction of the positive $x$-axis in S. Without any loss of generality,
  only points lying on the $x$,$x'$ axes are considered in the following. The epochs $t$,$t'$
   are those recorded by
   similar clocks at rest in S,S' respectively.
   \par The transformation equations (1) and (2) respect spatial and temporal translational
      invariance, that is they are unchanged by the replacements:
     \[ x \rightarrow x+X,~~~~ t \rightarrow t+T   \]
   where X and T are arbitary constants. They also remain invariant under the
    operation of space-time exchange (STE):
     \[ x \leftrightarrow x^0,~~~~ x' \leftrightarrow (x^0)'  \]
     which exchanges equations (1) and (2).  
      The STE invariance concept:
     \par {\tt The equations describing the laws of physics are
         invariant with respect to the exchange of space and time coordinates, or, more
     generally to the \newline exchange of the spatial and temporal components of four vectors.}
      \par was introduced in Ref.~\cite{JHFSTE}. A corollary is the independence of 
      physical predictions of any theory to the choice of metric (space-like or time-like)
    for four-vector
     products. As shown in  Ref.~\cite{JHFSTE} the postulate of STE invariance, together with
     the weak postulates of spatial homogeneity~\cite{LJE,YPT,JMLL} or single-valuedness~\cite{JHFHPA}
     is sufficient to derive the space-time LT (1) and (2). Another application of
     STE invariance is the derivation~\cite{JHFSTE}
 of the non-homogeneous electrodynamical (Amp\`{e}re's Law) and magnetodynamical (Faraday's
    Law of Induction) Maxwell equations from, respectively, the electrostatic and magnetostatic
    Gauss laws.
 \begin{figure}[htbp]
 \begin{center}\hspace*{-0.5cm}\mbox{
 \epsfysize10.0cm\epsffile{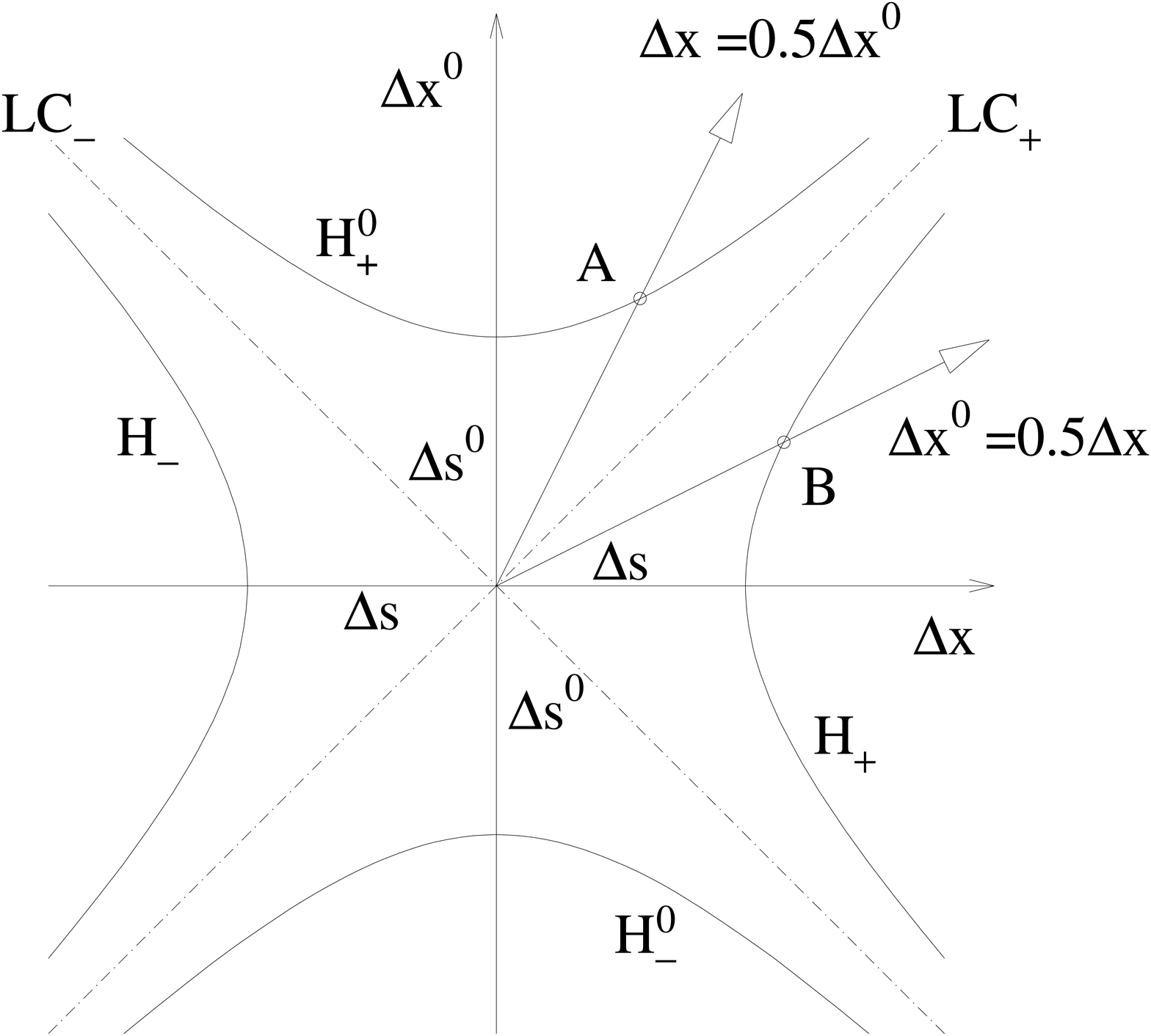}}
 \caption{{\em Minkowski $\Delta x^0$ versus $\Delta x$ plot. The STE conjugate worldlines
  $\Delta x = 0.5\Delta x^0$,$\Delta x^0 = 0.5\Delta x$ intersect the hyperbolae ${\rm H}^0_+$, ${\rm H}_+$
  corresponding, respectively, to time-like and space-like invariant interval relations, in the points
   A,B. See text for discussion.}}
 \label{fig-fig1}
 \end{center}
 \end{figure}
    \par The transformation equations (1) and (2) may be combined to define invariant interval
 equations as introduced by Minkowski~\cite{Minkowski}\footnote{See Ref.~\cite{JHFMinkplt}
  for a discussion of the consequences of a sign error in drawing the $x'$ and $t'$ axes on the
  original space-time plot of Ref.~\cite{Minkowski} as well as in a wide subsequent literature.} and discussed at length by 
   Langevin~\cite{Langevin,JHFL}:
      \begin{eqnarray}
    (\Delta x^0)^2 &-&(\Delta x)^2 = [\Delta(x^0)']^2-(\Delta x')^2 \equiv (\Delta s^0)^2~~
     {\rm Timelike~interval}~~\Delta x^0 > \Delta x,  \\
    (\Delta x)^2&-&(\Delta x^0)^2 = (\Delta x')^2- [\Delta(x^0)']^2 \equiv (\Delta s)^2~~
     {\rm Spacelike~interval}~~\Delta x > \Delta x^0.
      \end{eqnarray}
    For a fixed value of $\Delta s^0 = i\Delta s$ the intervals $\Delta x^0(\beta)$,
     $\Delta x(\beta)$ for different values of $\beta$ lie along four distinct hyperbolae
      ${\rm H}_+$, ${\rm H}_-$, ${\rm H}^0_+$ and  ${\rm H}^0_-$ in the $\Delta x^0$ versus $\Delta x$
    Minkowski plot, as shown in Fig. 1.
    The equations of the hyperbolae
      are:
    \begin{eqnarray}
    {\rm H}_+~:~~~\Delta x^0 & = &\pm\sqrt{(\Delta x)^2-(\Delta s)^2},~~~~\Delta x \ge\Delta s, \\
  {\rm H}_-~:~~~\Delta x^0 & = & \pm\sqrt{(\Delta x)^2-(\Delta s)^2},~~~~\Delta x \le -\Delta s, \\
 {\rm H}^0_+~~:~~~\Delta x & = & \pm\sqrt{(\Delta x^0)^2-(\Delta s^0)^2},~~~~\Delta x^0 \ge\Delta s^0, \\
  {\rm H}^0_-~~:~~~\Delta x & = & \pm\sqrt{(\Delta x^0)^2-(\Delta s^0)^2},~~~~\Delta x^0 \le -\Delta s^0. 
  \end{eqnarray}
   Since the physical significance of Fig. 1 does not depend on the direction in which the
   $\Delta x$ and $\Delta x^0$ axes are drawn, the figure is invariant under the STE operation.
   In fact, the sucessive operations STE, anticlockwise rotation by $90^{\circ}$ in the
   $\Delta x\Delta x^0$ plane and rotation by $180^{\circ}$ about the resulting $\Delta x^0$
   axis leave Fig. 1 unchanged. As will now be demonstrated, this manifest STE invariance is
   broken when the physical significance of various projection operators applied to the LT
   (1) and (2) is considered.
    \par Setting $\Delta x' = 0$ in Eq.~(1) means consideration of events on the world line 
     of a fixed point in the frame S'. The corresponding differential worldline equation
     in the frame S is, from Eq.~(1), $\Delta x = \beta \Delta x^0$. For $\beta = 0.5$ this
     straight line in Fig.~1 intersects the hyperbola ${\rm H}^0_+$ at the point A. Using
    the worldline equation in S to eliminate $\Delta x$ in Eq.~(2) yields the time dilation
    relation $\Delta x^0 = \gamma \Delta(x^0)'$ which is the
    experimentally-confirmed~\cite{NatureTD,AshbyPT}
    prediction that clocks at rest in the frame S' are seen to
    run slow relative to clocks at rest in the frame S. 
     \par The STE conjugate projection  $\Delta (x^0)' = 0$, i.e. simultaneous events 
     in the frame S', gives from Eq.~(2) the relation  $\Delta x = \Delta x^0/\beta $
     corresponding to a superlumial worldline in the frame S that intersects the hyperbola ${\rm H}_+$
     in Fig.~1 at the point B for $\beta = 0.5$. Since for $\Delta x > 0$ and $\beta > 0$
     then also $\Delta x^0 > 0$, there is here a `relativity of simultaneity' effect because
     events simultaneous in S' ($\Delta (x^0)' = 0$) are not so in the frame S ($\Delta x^0 > 0$).
     Using the worldline equation to eliminate $\Delta x^0$ in Eq.~(1) gives
     $\Delta x = \gamma \Delta x'$. This is the `space dilation' effect (the STE conjugate
     of time dilation) associated with the
     projection  $\Delta (x^0)' = 0$ as previously pointed out in Ref.~\cite{JHFTNRSE}. However,
    {\it any object} at rest in the frame S' must have $\Delta x' = 0$. So it is impossible
     that the worldline of any physical clock at rest in S' can intersect the hyperbola ${\rm H}_+$.
     The mathematical projection  $\Delta (x^0)' = 0$ with its associated relativity of 
     simultaneity and `space dilation' effects is therefore unphysical. The initial 
     conditions of an experiment where a clock a rest in S' is compared with one at rest in S:
      $\beta > 0$, $\Delta x' = 0$ therefore restrict the physical region of the
      Minkowski plot in Fig.~1 to one eighth of its total area ---that between the lightcone
      LC$_+$ (the asymptote of the hyperbola ${\rm H}^0_+$) and
     the positive $\Delta x^0$ axis.  The symmetry of the plot is therefore clearly broken
     by the initial conditions of any experiment in which the time dilation effect is observed.
 \begin{figure}[htbp]
 \begin{center}\hspace*{-0.5cm}\mbox{
 \epsfysize10.0cm\epsffile{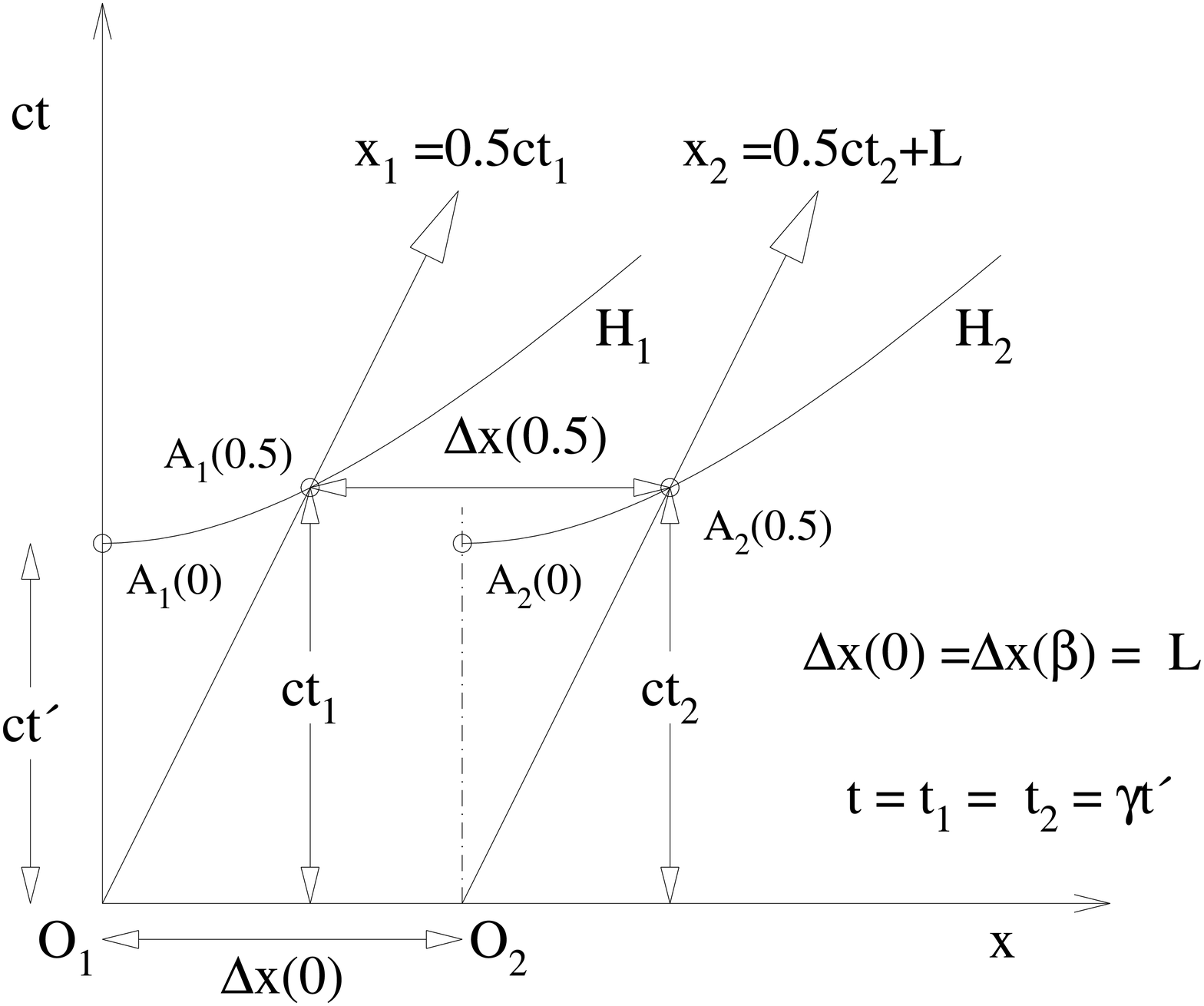}}
 \caption{{\em Minkowski $ct$ versus $x$ plot for clocks ${\rm C}_1$ and ${\rm C}_2$ at rest in the
  frame S' with respective worldlines: $x_1 = \beta c t_1$ and $x_2 = \beta c t_2+L$ in the 
   frame S. The hyperbolae ${\rm H}_1$ and ${\rm H}_2$ are the loci of points (x,ct) for a fixed 
   value of $t'$ and different positive values of $\beta$. O$_1$A$_1(0.5)$, O$_2$A$_2(0.5)$: worldlines
   in S of ${\rm C}_1$ and ${\rm C}_2$  for $\beta = 0.5$. O$_1$A$_1(0)$, O$_2$A$_2(0)$: similar
   wordlines for $\beta = 0$. The absence of any `relativity of simultaneity' or `length contraction'
  effects is evident from inspection of this figure.}}
 \label{fig-fig2}
 \end{center}
 \end{figure}
     \par Consider now two clocks ${\rm C}_1$ and ${\rm C}_2$ at rest in S' with worldlines
    in the frame S: $x_1(\beta) = \beta c t_1$ and  $x_2(\beta) = \beta c t_2+L$. Integrating
    the differential time dilation relation $dt = \gamma dt'$ gives time dilation relations
    $t_1 = \gamma t'_1$,  $t_2 = \gamma t'_2$ for  ${\rm C}_1$, ${\rm C}_2$. Use of the identity
    $\gamma^2(1-\beta^2)\equiv 1$ then yields the relations:
    \begin{eqnarray}
   c^2(t_1)^2&-&x_1(\beta)^2 = c^2(t'_1)^2, \\
    c^2(t_2)^2&-&(x_2(\beta)-L)^2 = c^2(t'_2)^2.
    \end{eqnarray}
  The corresponding hyperbolae ${\rm H}_1$ and ${\rm H}_2$ on a $ct$ versus $x$ plot are
  shown in Fig.~2 for $t'_1 = t'_2 =  t'$, together with the worldlines of ${\rm C}_1$
  and ${\rm C}_2$ for $\beta = 0.5$ and $\beta = 0$. It follows from the time dilation relations or
  inspection of Fig.~2 that $t_1 = t_2$ when $t'_1 = t'_2$ ---there is no `relativity
  of simultaneity' effect in observations of the clocks  ${\rm C}_1$ and ${\rm C}_2$. 
  The worldline equations when  $t_1 = t_2$ show that, for all values
  of $\beta$:
  \begin{equation}
  \Delta x(\beta) \equiv x_2(\beta) - x_1(\beta) = L.
  \end{equation}
   A special case of Eq.~(12) is
   \begin{equation}
   x_2(0) - x_1(0) = x'_2- x'_1 \equiv \Delta x' =
     x_2(\beta) - x_1(\beta) \equiv \Delta x(\beta) = L
   \end{equation} 
    so, as is also evident from inspection of Fig.~2, there is no `length contraction'
   effect.
   \par For further discussion of the invariance of measured length intervals ---a property which
     is independent of the form of space-time transformation equations--- see Ref.~\cite{JHFFJMP2}.
   \par The interval LT (1) and (2) for the clock ${\rm C}_2$ are:
      \begin{eqnarray}
     x'_2-L & = & \gamma(x_2-L-\beta ct_2) = 0, \\
    ct'_2 & = & \gamma[ct_2 -\beta(x_2-L)]. 
   \end{eqnarray}
    The corresponding LT for clock ${\rm C}_1$ are given by setting $L=0$ in these equations.
    \par It is now instructive to compare (13) and (14) with the conventional space-time LT~\cite{Ein1}:
       \begin{eqnarray}
     x' & = & \gamma(x-vt), \\
    t' & = & \gamma\left(t -\frac{vx}{c^2}\right) 
   \end{eqnarray}    
   which has hitherto been universally interpreted as the transformation 
   giving the observed event in the frame S': ($x'$,$t'$) corresponding to an event($x$,$t$)
   observed in the frame S, for {\it arbitary values} of $x'$,$t'$ or $x$,$t$. However, since the
   coordinate $x'$ is, by definition, that of a fixed point in the frame S' it must be 
   independent of time. In contrast, the right side of Eq.~(15) is in general a function of the time
   $t$ which, for any value of $x$, vanishes when $t= x/v$. It then necessarily follows that
   Eq.~(15) {\it can hold only if both $x' =0$ and $x = vt$}, in which case (15) and (16) 
   become identical to (13) and (14) with $L =0$, i.e. the correct interval LT for the
   clock ${\rm C}_1$ discussed above. 
   \par The spurious `length contraction' and `relativity of simultaneity' effects derived from (15) and (16),
     discussed in detail elesewhere~\cite{JHFSTP3,JHFRECP,JHFLSTP}, arise from the failure to respect
     the above-mentioned condition for the validity of Eq.~(15). The LT (15) and (16)
     are instead assumed to hold for arbitary values of $x'$, so they become, on considering 
     two independent events:
     \begin{eqnarray} 
       x'_1 & = & \gamma(x_1-vt_1),~~~~ x'_2  =  \gamma(x_2-vt_2), \\
  t'_1 & = & \gamma\left(t_1 -\frac{vx_1}{c^2}\right),
   ~~~~t'_2 = \gamma\left(t_2 -\frac{vx_2}{c^2}\right). 
   \end{eqnarray}    
    On setting $t_1 = t_2$ ($\Delta t = 0$, length measurement in the frame S) Eqs.~(17) give:
   \[   x'_2- x'_1 \equiv \Delta x' = \gamma(x_2-x_1) \equiv \gamma \Delta x
      ~~({\rm length~contraction}) \]
    while Eqs.~(18) give:
   \[t'_2- t'_1\equiv \Delta t' = -\frac{\gamma v(x_2-x_1)}{c^2} = -\frac{\gamma  \Delta x'}{c^2}
     \ne 0~~({\rm relativity~of~simultaneity}). \]
    These unphysical predictions therefore arise from a failure to sufficiently consider the
    mathematical constraints arising from the
    operational meanings of the coordinate symbols in the LT. 
     \par The erroneous (when $x' \ne 0$) LT equations (15) and (16) differ from the correct
    ones (13) and (14) by the omission of certain additive constants $X$ and $T$ on the right side
     of (15) and (16) respectively. As discussed in Ref.~\cite{JHFFJMP1} the necessity to
     include such constants to correctly describe synchronised clocks at different 
     spatial positions was clearly pointed out by Einstein in Ref.~\cite{Ein1} 
     but, to the present author's best knowledge, was never done, either by him
     or any subsequent worker, for the entire duration of the 20th Century!
     \par The physical meaning of Eqs.~(13) and (14) is the same as that of the more
     transparent equations:
    \begin{eqnarray}  
    x'_2 = L, & &~~~~~x_2 = vt_2+L, \\
     t_2 & = & \gamma t'_2. 
   \end{eqnarray}
     The first and second equations in (19) are simply the
    worldline equations of ${\rm C}_2$ in the
   frames S', S respectively and are the same as in Galilean relativity. The only modification
   of space-time transformation equations in passing from Galilean to special relativity is
   the replacement of Newtonian universal time: $T = t = t'$ by the position-independent
   time dilation relation (20).
   \par Note that, as is also evident by inspection of Fig.~2, the worldlines of
    ${\rm C}_1$ and ${\rm C}_2$ in the frames S', S respectively, respect, at any instant:
     $t = t_1 = t_2$ translational invariance: $x'_2 = x'_1+L$,  $x_2 = x_1+L$, as do the
     interval transformations for events on the worldlines of the clocks ${\rm C}_i$, $i=1,2$:
    \begin{eqnarray}  
   \Delta x'_i = 0, & &~~~~~ \Delta x_i = \beta \Delta x^0_i, \\
     \Delta x^0_i & = & \gamma   \Delta (x^0_i)'. 
   \end{eqnarray}
    Comparing Eqs.~(21) and (22) with the general, STE invariant, interval Lorentz transformations
    (1) and (2), the breakdown of STE invariance in space-time experiments involving such
     physical clocks is manifest.

\end{document}